\documentclass[aps,pra,preprint,superscriptaddress]{revtex4-1}

\usepackage{amsmath,amsfonts,amssymb}
\usepackage{xcolor,graphicx}
\usepackage{braket}
\usepackage{float}
\usepackage[pdftex]{hyperref}
\usepackage{csquotes}
\hypersetup{
	colorlinks = true,
	linkcolor = blue,
	anchorcolor = blue,
	citecolor = red,
	filecolor = blue,
	urlcolor = blue}

\begin{document}

\title{Paraxial wave propagation: Operator techniques }

\author{H.~M.~Moya-Cessa}
\email[e-mail: ]{hmmc@inaoep.mx}
\affiliation{Instituto Nacional de Astrofísica Óptica y Electrónica (INAOE)\\ Luis Enrique Erro 1, Santa María Tonantzintla, Puebla, 72840, Mexico}
\author{I.~Ramos-Prieto}
\affiliation{Instituto Nacional de Astrofísica Óptica y Electrónica (INAOE)\\ Luis Enrique Erro 1, Santa María Tonantzintla, Puebla, 72840, Mexico}
\author{F.~Soto-Eguibar}
\affiliation{Instituto Nacional de Astrofísica Óptica y Electrónica (INAOE)\\ Luis Enrique Erro 1, Santa María Tonantzintla, Puebla, 72840, Mexico}
\author{U.~Ruíz}
\affiliation{Instituto Nacional de Astrofísica Óptica y Electrónica (INAOE)\\ Luis Enrique Erro 1, Santa María Tonantzintla, Puebla, 72840, Mexico}
\author{D.~{Sánchez}-{de la Llave}}
\affiliation{Instituto Nacional de Astrofísica Óptica y Electrónica (INAOE)\\ Luis Enrique Erro 1, Santa María Tonantzintla, Puebla, 72840, Mexico}

\begin{abstract}
    The similarity between the Schrödinger equation and 
    the paraxial wave equation permits numerous analogies linking these fields, which is pivotal in advancing both quantum mechanics and wave optics. In this study, we demonstrate the application of operator techniques to an electromagnetic field characterized by the function $f(x + ay)$, leveraging the structural analogies between these equations. Specifically, we employ initial conditions defined by Airy and Bessel functions to illustrate the practical implementation of these techniques.
\end{abstract}

\date{\today}

\maketitle

\section{Introduction}
The propagation of light beams is a crucial topic in both applied and theoretical optics, with significant attention given over the years to beams that can maintain their intensity profiles during propagation. These beams, known as adiffractional beams, are of particular interest because their transverse intensity profiles remain unchanged, without scaling, as they propagate. Notable examples include Mathieu beams~\cite{Gutierrez_2000}, Bessel beams~\cite{Durnin_1987, McGloin_2005, Fahrbach_2010, Gori_1987}, and Weber beams~\cite{Bandres_2004}, all of which exhibit this remarkable feature. These beams have numerous practical applications, ranging from optical trapping and manipulation to advanced imaging techniques~\cite{Ashkin_1987}. The experimental implementation of these beams necessitates the inclusion of a finite envelope~\cite{Gori_1987, Khonina_1999, Siviloglou_2007, Arturo, Bandres_2007, Khonina_2011}, thereby limiting their propagation invariance to a finite region. Additionally, there are optical fields that can preserve their shape over a certain propagation range, known as scaled propagation invariant beams~\cite{Arrizon_2018}. However, this property is achieved at the expense of a re-scaling in their transverse dimensions.

In this contribution, we employ operator techniques~\cite{Stoler_1980, David, Libro} to investigate the propagation characteristics of a field described by the function $f(x+ay)$. Specifically, our focus extends to the case where $a=i$ (with $i = \sqrt{-1}$), a scenario previously explored in~\cite{CRBI, CRBII}. By leveraging these operator methods, we elucidate the dynamics of such fields during propagation. In particular we will show that when the function is an Airy function, which has the well-known feature that it bends during propagation \cite{Siviloglou_2007,Arturo} such bending may be accelerated for $a\ne 0$. 

\section{Paraxial propagation}
The propagation of light in free space is governed by the paraxial wave equation:
\begin{equation}
ik\frac{\partial E(x,y,z)}{\partial z} = -\frac{1}{2}\left(\frac{\partial^2 E(x,y,z)}{\partial x^2} + \frac{\partial^2 E(x,y,z)}{\partial y^2}\right),
\end{equation}
where $k$ is the wavevector, defined as $k = \frac{2\pi}{\lambda}$. To facilitate the use of operator notation, we introduce the momentum operators $\hat{p}_{\alpha} = -i\frac{\partial}{\partial \alpha}$, where $\alpha$ represents the transverse coordinates $x$ and $y$. These operators are analogous to the canonical momentum operators in quantum mechanics. With these definitions, the paraxial wave equation can be rewritten in a more compact form:
\begin{equation}\label{Paraxial2}
i\frac{\partial E(z)}{\partial z} = \frac{\hat{p}_x^2 + \hat{p}_y^2}{2k} E(z).
\end{equation}
This form highlights the analogy between the paraxial wave equation and the Schrödinger equation for a free particle, where the propagation coordinate $z$ plays the role of time, and the transverse momentum operators $\hat{p}_x$ and $\hat{p}_y$ correspond to the kinetic energy terms. The formal solution to Eq.~(\ref{Paraxial2}) can be expressed as the exponential of the right-hand side of this equation, representing the free particle evolution operator. Specifically, the solution is given by:
\begin{equation}\label{0050}
E(x,y,z) = \exp\left[-i\frac{z}{2k} \left(\hat{p}_x^2 + \hat{p}_y^2 \right) \right] E(x,y,0),
\end{equation}
where $E(x,y,0)$ represents the initial field distribution at $z = 0$. This solution is derived by recognizing that the operator $\left(\hat{p}_x^2 + \hat{p}_y^2\right)$ generates free-space propagation in the transverse plane. The exponential operator in Eq.~(\ref{0050}) can be interpreted as a propagator that evolves the field $E(x,y,0)$ over a distance $z$. This formalism is particularly useful for studying the propagation of optical beams and pulses, allowing for the analysis of diffraction and focusing effects in a systematic manner. 

Now, our focus turns to the initial condition. We consider the initial field $E(x,y,0) = f(x + ay)$, which allows us to express the solution as:
\begin{equation}
\begin{split}
E(x,y,z) &= \exp\left[-i\frac{z}{2k} \left(\hat{p}_x^2 + \hat{p}_y^2 \right) \right] f(x + ay)\\
         &= \exp(-iay\hat{p}_x)\exp(iay\hat{p}_x)\exp\left[-i\frac{z}{2k} \left(\hat{p}_x^2 + \hat{p}_y^2 \right) \right] \exp(-iay\hat{p}_x) f(x),
\end{split}
\end{equation}
where we have used the commutation relation $[\alpha,\hat{p}_\alpha] = i$ (with $\alpha = x,y$), implying $f(x + ay) = \exp(-iay\hat{p}_x)f(x)$. Consequently, by applying a bilateral relation to the operator $\exp\left[-i\frac{z}{2k} \left(\hat{p}_x^2 + \hat{p}_y^2 \right) \right]$, we obtain
\begin{equation}
\begin{split}
E(x,y,z) &= \exp(-iay\hat{p}_x)\exp\left\{-i\frac{z}{2k} \left[\hat{p}_x^2 + (\hat{p}_y + a\hat{p}_x)^2 \right] \right\} f(x)\\
&= \exp(-iay\hat{p}_x) \exp\left\{-i\frac{z}{2k}\left[(1 + a^2)\hat{p}_x^2 + \hat{p}_y^2 + 2a\hat{p}_x\hat{p}_y \right] \right\} f(x).
\end{split}
\end{equation}
It is noteworthy that both exponentials containing the operator $\hat{p}_y$ applied to the function $f(x)$ leave the function unchanged. Thus, we arrive at the solution:
\begin{equation}\label{Campo}
    E(x,y,z) = \exp(-iay\hat{p}_x)\exp\left[-i\frac{(1+a^2)z}{2k} \hat{p}_x^2\right]f(x).
\end{equation}
This equation reveals that the propagation is scaled by $1+a^2$, indicating faster effects (as discussed further in the next section where we explore the acceleration of Airy beams). Another notable feature illustrated here is that when $a = i$, the function $f(x + iy)$ becomes diffraction-free~\cite{CRBI}.

\section{Airy function} 
As an illustrative and representative example, we consider the case where the initial field ${E}(x,y,0)$ is described by the Airy function:
\begin{equation}\label{InitialA}
{E}(x,y,0) = \mathrm{Ai}\left[\eta (x + ay)\right],
\end{equation}
where $\eta$ is a scaling parameter. Berry and Balazs~\cite{Berry} demonstrated that the free evolution of such a wave packet (for $a=0$ and $k=1$) leads to the expression:
\begin{equation}\label{AiryPropagated}
\exp\left(-i\frac{z}{2}\hat{p}_x^2\right)\mathrm{Ai}(\eta x) = \exp\left[i\frac{\eta^3 z}{2} \left(x - \frac{\eta^3 z^2}{6}\right)\right] \mathrm{Ai}\left[\eta \left(x - \frac{\eta^3 z^2}{4}\right)\right].
\end{equation}
Building on this result, by considering both Eq.~(\ref{Campo}) and Eq.~(\ref{InitialA}), and redefining the longitudinal coordinate as $z \rightarrow z(1 + a^2)$, we derive the expression for the propagated optical field. The resulting field $E(x,y,z)$ is given by:
\begin{equation}\label{Airy1}
    \begin{split}
E(x,y,z) &= \exp\left[i\frac{\eta^3 z (1 + a^2)}{2} \left(x + ay - \frac{\eta^3 z^2 (1 + a^2)^2}{6}\right)\right]\\&\times\mathrm{Ai}\left[\eta \left(x + ay - \frac{\eta^3 z^2 (1 + a^2)^2}{4}\right)\right].
    \end{split}
\end{equation}
Furthermore, for comparative purposes, we analyze the product of two Airy functions. From Eq.~(\ref{AiryPropagated}), we can derive that under this initial condition, the field is determined by
\begin{equation}\label{Airy2}
    \begin{split}
\mathcal{E}(x,y,z) &= \exp\left[i\frac{\eta^3 z}{2} \left(x - \frac{\eta^3 z^2}{6}\right)\right] \exp\left[i\frac{\eta^3 z}{2} \left(y - \frac{\eta^3 z^2}{6}\right)\right]\\&\times \mathrm{Ai}\left[\eta \left(x - \frac{\eta^3 z^2}{4}\right)\right] \mathrm{Ai}\left[\eta \left(y - \frac{\eta^3 z^2}{4}\right)\right].
    \end{split}
\end{equation}

\begin{figure}
\centering
\includegraphics[width=.75\linewidth]{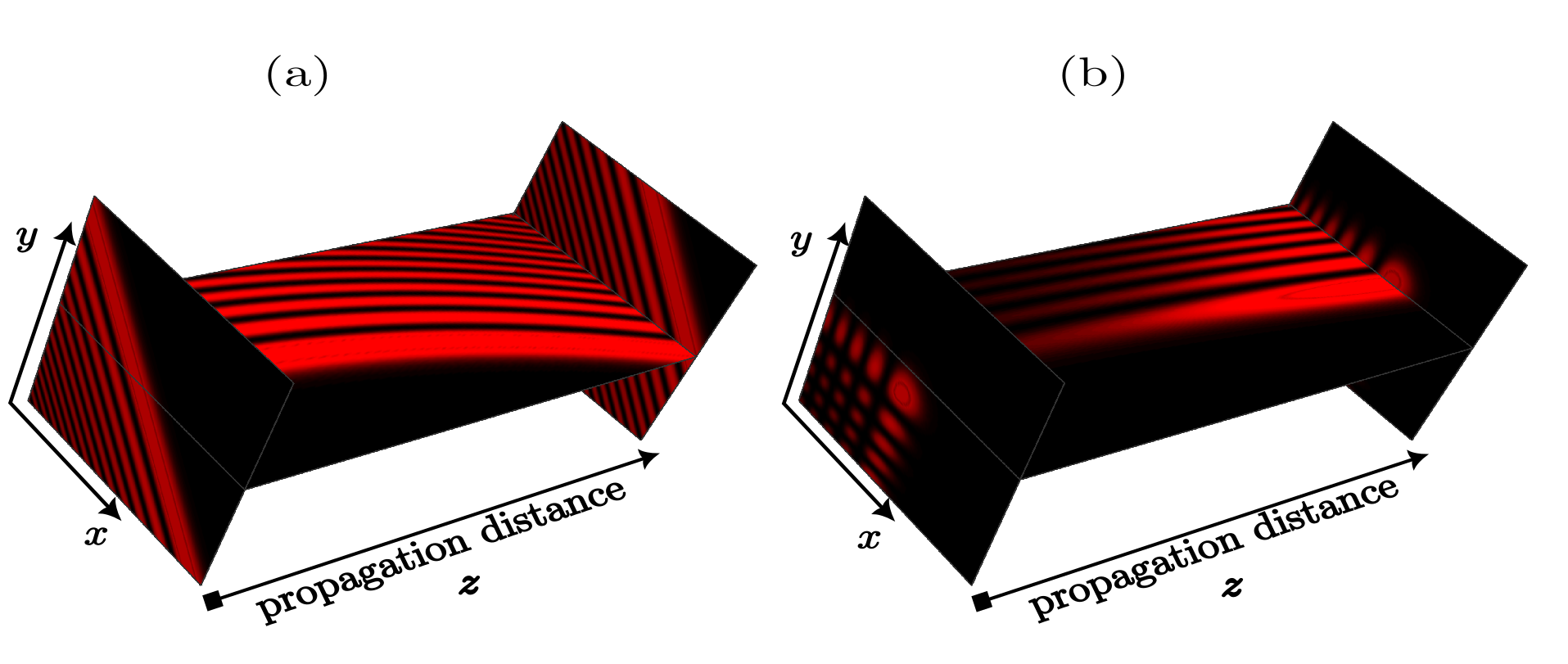}
\caption{(a)~Given the initial condition at $z=0$ from Eq.~(\ref{Airy1}), we can represent the propagation at different distances. Specifically, we consider the case of $a=1$ to illustrate that this parameter accelerates the bending along the plane, for example, $y=0$ (without loss of generality, we choose this plane, but similar behavior occurs in any plane along the propagation direction). (b)~Using Eq.~(\ref{Airy2}), we observe a bending in the beam; however, although the propagation distances are the same as in the previous case, the bending is less noticeable due to the initial condition (the product of Airy functions). The parameters used in both cases are: $a=1$, planes at $z=\left\{0,5\right\}$ (dimensionless units), and $\eta=0.75$.
}
\label{fig1}
\end{figure}
In Fig.~\ref{fig1}, we present the propagation dynamics of the Airy function as described by Eqs.~(\ref{Airy1}) and (\ref{Airy2}). The plots clearly demonstrate how the beam's bending occurs earlier when the parameter $a \ne 0$, effectively accelerating the bending effect. This early onset of the bending is a direct consequence of the transverse coordinate modulation introduced by the parameter $a$. The visual representation in Fig.~\ref{fig1} highlights the significant influence of this parameter on the beam's trajectory, showcasing the ability to control and manipulate the propagation characteristics of the Airy beam through appropriate parameter selection. 

\section{Bessel function}
\begin{figure}
\centering
\includegraphics[width=.75\linewidth]{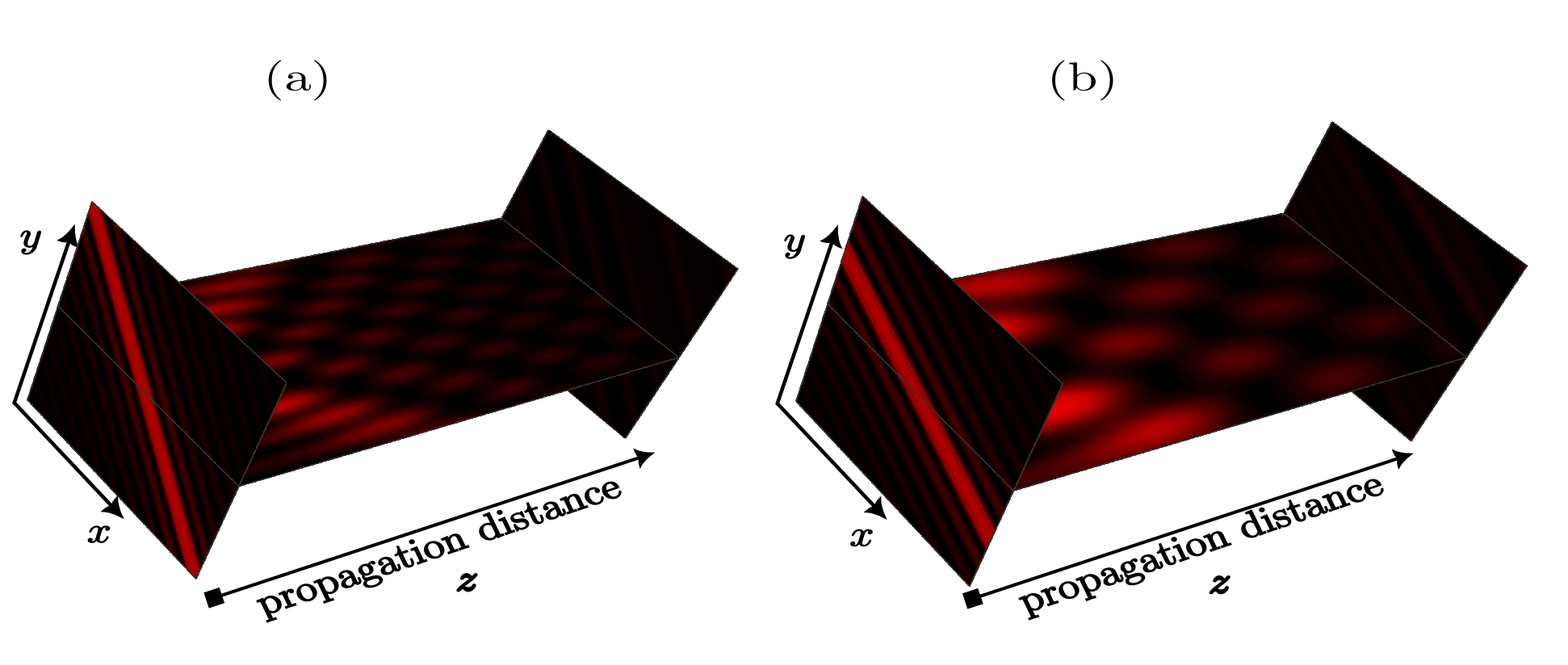}
\caption{From Eq.~(\ref{Bessel}), we plot the field in the planes $z=\left\{0, 30\right\}$ (dimensionless units) and in the plane $x=0$, which shows the propagation of the initial condition in that plane, with $\eta=1$ and $n=0$. In (a), we present the case for $a=1$, and in (b), the case for $a=0.75$. For these two values of $a$, we can observe how this parameter controls the transverse inclination of $J_n[\eta(x+ay)]$, and specifically, it results in a shift in the direction of propagation.}
\label{fig2}
\end{figure}

We analyze the propagation of a Bessel function of the first kind, namely
\begin{equation}\label{InitialBessel}
    E(x,y,0)=J_n(\eta [x+ay]),
\end{equation}
where $J_n$ is the Bessel function of the first kind, $n$ is the order of the Bessel function, $\eta$ is a scaling parameter. To analyze the evolution of this initial field as it propagates, we apply the paraxial approximation of the Helmholtz equation. It can be shown that the action of the propagation operator $\exp\left(-i\frac{z}{2}\hat{p}x^2\right)$ on the Bessel function $J_n\left(\eta x\right)$ leads to the following expression \cite{Optica}~(with $k=1$):
\begin{equation}
\exp\left(-i\frac{z}{2}\hat{p}x^2\right) J_n\left(\eta x\right) = \exp\left(-i\frac{z\eta^2}{4}\right)
\sum{l=-\infty}^{\infty} i^l J{n+2l}(\eta x) J_l\left(\frac{z\eta^2}{4}\right).
\end{equation}
The above sum is recognized as a generalized Bessel function \cite{Dattoli}
\begin{equation}
\mathcal{J}_n(x,y;s)=\sum_{l=-\infty}^{\infty}(-s)^lJ_{n+2l}( x)J_l\left(y\right).
\end{equation}
Then the propagated field is
\begin{equation}\label{Bessel} 
E(x,y,z) =\exp\left[-i\frac{z(1+a^2)\eta^2}{4}\right] 
\sum_{l=-\infty}^{\infty}i^lJ_{n+2l}\left[\eta (x+ay)\right]J_l\left[\frac{z(1+a^2)\eta^2}{4}\right].
\end{equation}
It is clear from this expression that for $z=0$ the only Bessel function that is non zero is the zeroth order one so we recover the {\it initial} function Eq.~(\ref{InitialBessel}). On the other hand, it may also be seen from the above equation that it is a clear cut of the propagated field at $x=-ay$. For such values the field is always zero independent of the value of the propagation distance. This behavior is illustrated for two different values of $a=\left\{1, 0.5\right\}$ in Fig.~\ref{fig2}.

\section{Conclusions}
It has been shown how to propagate a field of the form $f(x+ay)$ using operator techniques. The solution demonstrates an acceleration in propagation that depends on the parameter $a$. This acceleration is clearly illustrated by propagating an Airy function of $x+ay$ and comparing it with a normal Airy function of $x$; the bending becomes stronger with increasing values of the parameter $a$.

For Bessel functions, specifically $J_n(x+ay)$, it has been demonstrated that when the order $n$ is non-zero, the field is zero precisely at $x=-ay$. Therefore, the parameter $a$ controls the spatial point at which the field vanishes. This observation underscores how $a$ influences the spatial distribution of the propagated field, particularly in comparison with the behavior of standard Bessel functions at $x$.

\begin{acknowledgements}
    H.M. M.-C. acknowledges the invitation to deliver a talk in the congress to honour Professor David J. Fern\'andez's 40 years of research with whom he coauthored the paper in Reference \cite{David}.
\end{acknowledgements}

\begin{thebibliography}{21}%
    \makeatletter
    \providecommand \@ifxundefined [1]{%
     \@ifx{#1\undefined}
    }%
    \providecommand \@ifnum [1]{%
     \ifnum #1\expandafter \@firstoftwo
     \else \expandafter \@secondoftwo
     \fi
    }%
    \providecommand \@ifx [1]{%
     \ifx #1\expandafter \@firstoftwo
     \else \expandafter \@secondoftwo
     \fi
    }%
    \providecommand \natexlab [1]{#1}%
    \providecommand \enquote  [1]{``#1''}%
    \providecommand \bibnamefont  [1]{#1}%
    \providecommand \bibfnamefont [1]{#1}%
    \providecommand \citenamefont [1]{#1}%
    \providecommand \href@noop [0]{\@secondoftwo}%
    \providecommand \href [0]{\begingroup \@sanitize@url \@href}%
    \providecommand \@href[1]{\@@startlink{#1}\@@href}%
    \providecommand \@@href[1]{\endgroup#1\@@endlink}%
    \providecommand \@sanitize@url [0]{\catcode `\\12\catcode `\$12\catcode `\&12\catcode `\#12\catcode `\^12\catcode `\_12\catcode `\%12\relax}%
    \providecommand \@@startlink[1]{}%
    \providecommand \@@endlink[0]{}%
    \providecommand \url  [0]{\begingroup\@sanitize@url \@url }%
    \providecommand \@url [1]{\endgroup\@href {#1}{\urlprefix }}%
    \providecommand \urlprefix  [0]{URL }%
    \providecommand \Eprint [0]{\href }%
    \providecommand \doibase [0]{http://dx.doi.org/}%
    \providecommand \selectlanguage [0]{\@gobble}%
    \providecommand \bibinfo  [0]{\@secondoftwo}%
    \providecommand \bibfield  [0]{\@secondoftwo}%
    \providecommand \translation [1]{[#1]}%
    \providecommand \BibitemOpen [0]{}%
    \providecommand \bibitemStop [0]{}%
    \providecommand \bibitemNoStop [0]{.\EOS\space}%
    \providecommand \EOS [0]{\spacefactor3000\relax}%
    \providecommand \BibitemShut  [1]{\csname bibitem#1\endcsname}%
    \let\auto@bib@innerbib\@empty
    \bibitem [{\citenamefont {Guti\'{e}rrez-Vega}\ \emph {et~al.}(2000)\citenamefont {Guti\'{e}rrez-Vega}, \citenamefont {Iturbe-Castillo},\ and\ \citenamefont {Ch\'{a}vez-Cerda}}]{Gutierrez_2000}%
      \BibitemOpen
      \bibfield  {author} {\bibinfo {author} {\bibfnamefont {J.~C.}\ \bibnamefont {Guti\'{e}rrez-Vega}}, \bibinfo {author} {\bibfnamefont {M.~D.}\ \bibnamefont {Iturbe-Castillo}}, \ and\ \bibinfo {author} {\bibfnamefont {S.}~\bibnamefont {Ch\'{a}vez-Cerda}},\ }\href {\doibase 10.1364/OL.25.001493} {\bibfield  {journal} {\bibinfo  {journal} {Opt. Lett.}\ }\textbf {\bibinfo {volume} {25}},\ \bibinfo {pages} {1493} (\bibinfo {year} {2000})}\BibitemShut {NoStop}%
    \bibitem [{\citenamefont {Durnin}\ \emph {et~al.}(1987)\citenamefont {Durnin}, \citenamefont {Miceli},\ and\ \citenamefont {Eberly}}]{Durnin_1987}%
      \BibitemOpen
      \bibfield  {author} {\bibinfo {author} {\bibfnamefont {J.}~\bibnamefont {Durnin}}, \bibinfo {author} {\bibfnamefont {J.~J.}\ \bibnamefont {Miceli}}, \ and\ \bibinfo {author} {\bibfnamefont {J.~H.}\ \bibnamefont {Eberly}},\ }\href {\doibase 10.1103/PhysRevLett.58.1499} {\bibfield  {journal} {\bibinfo  {journal} {Phys. Rev. Lett.}\ }\textbf {\bibinfo {volume} {58}},\ \bibinfo {pages} {1499} (\bibinfo {year} {1987})}\BibitemShut {NoStop}%
    \bibitem [{\citenamefont {McGloin}\ and\ \citenamefont {Dholakia}(2005)}]{McGloin_2005}%
      \BibitemOpen
      \bibfield  {author} {\bibinfo {author} {\bibfnamefont {D.}~\bibnamefont {McGloin}}\ and\ \bibinfo {author} {\bibfnamefont {K.}~\bibnamefont {Dholakia}},\ }\href {\doibase 10.1080/0010751042000275259} {\bibfield  {journal} {\bibinfo  {journal} {Contemporary Physics}\ }\textbf {\bibinfo {volume} {46}},\ \bibinfo {pages} {15} (\bibinfo {year} {2005})},\ \Eprint {http://arxiv.org/abs/https://doi.org/10.1080/0010751042000275259} {https://doi.org/10.1080/0010751042000275259} \BibitemShut {NoStop}%
    \bibitem [{\citenamefont {Fahrbach}\ \emph {et~al.}(2010)\citenamefont {Fahrbach}, \citenamefont {Simon},\ and\ \citenamefont {Rohrbach}}]{Fahrbach_2010}%
      \BibitemOpen
      \bibfield  {author} {\bibinfo {author} {\bibfnamefont {F.~O.}\ \bibnamefont {Fahrbach}}, \bibinfo {author} {\bibfnamefont {P.}~\bibnamefont {Simon}}, \ and\ \bibinfo {author} {\bibfnamefont {A.}~\bibnamefont {Rohrbach}},\ }\href {\doibase 10.1038/nphoton.2010.204} {\bibfield  {journal} {\bibinfo  {journal} {Nature Photonics}\ }\textbf {\bibinfo {volume} {4}},\ \bibinfo {pages} {780} (\bibinfo {year} {2010})}\BibitemShut {NoStop}%
    \bibitem [{\citenamefont {Gori}\ \emph {et~al.}(1987)\citenamefont {Gori}, \citenamefont {Guattari},\ and\ \citenamefont {Padovani}}]{Gori_1987}%
      \BibitemOpen
      \bibfield  {author} {\bibinfo {author} {\bibfnamefont {F.}~\bibnamefont {Gori}}, \bibinfo {author} {\bibfnamefont {G.}~\bibnamefont {Guattari}}, \ and\ \bibinfo {author} {\bibfnamefont {C.}~\bibnamefont {Padovani}},\ }\href {\doibase https://doi.org/10.1016/0030-4018(87)90276-8} {\bibfield  {journal} {\bibinfo  {journal} {Optics Communications}\ }\textbf {\bibinfo {volume} {64}},\ \bibinfo {pages} {491} (\bibinfo {year} {1987})}\BibitemShut {NoStop}%
    \bibitem [{\citenamefont {Bandres}\ \emph {et~al.}(2004)\citenamefont {Bandres}, \citenamefont {Guti\'{e}rrez-Vega},\ and\ \citenamefont {Ch\'{a}vez-Cerda}}]{Bandres_2004}%
      \BibitemOpen
      \bibfield  {author} {\bibinfo {author} {\bibfnamefont {M.~A.}\ \bibnamefont {Bandres}}, \bibinfo {author} {\bibfnamefont {J.~C.}\ \bibnamefont {Guti\'{e}rrez-Vega}}, \ and\ \bibinfo {author} {\bibfnamefont {S.}~\bibnamefont {Ch\'{a}vez-Cerda}},\ }\href {\doibase 10.1364/OL.29.000044} {\bibfield  {journal} {\bibinfo  {journal} {Opt. Lett.}\ }\textbf {\bibinfo {volume} {29}},\ \bibinfo {pages} {44} (\bibinfo {year} {2004})}\BibitemShut {NoStop}%
    \bibitem [{\citenamefont {Ashkin}\ \emph {et~al.}(1987)\citenamefont {Ashkin}, \citenamefont {Dziedzic},\ and\ \citenamefont {Yamane}}]{Ashkin_1987}%
      \BibitemOpen
      \bibfield  {author} {\bibinfo {author} {\bibfnamefont {A.}~\bibnamefont {Ashkin}}, \bibinfo {author} {\bibfnamefont {J.~M.}\ \bibnamefont {Dziedzic}}, \ and\ \bibinfo {author} {\bibfnamefont {T.}~\bibnamefont {Yamane}},\ }\href {\doibase 10.1038/330769a0} {\bibfield  {journal} {\bibinfo  {journal} {Nature}\ }\textbf {\bibinfo {volume} {330}},\ \bibinfo {pages} {769} (\bibinfo {year} {1987})}\BibitemShut {NoStop}%
    \bibitem [{\citenamefont {Khonina}\ \emph {et~al.}(1999)\citenamefont {Khonina}, \citenamefont {Kotlyar}, \citenamefont {Soifer}, \citenamefont {Lautanen}, \citenamefont {Honkanen},\ and\ \citenamefont {Turunen}}]{Khonina_1999}%
      \BibitemOpen
      \bibfield  {author} {\bibinfo {author} {\bibfnamefont {S.}~\bibnamefont {Khonina}}, \bibinfo {author} {\bibfnamefont {V.}~\bibnamefont {Kotlyar}}, \bibinfo {author} {\bibfnamefont {V.}~\bibnamefont {Soifer}}, \bibinfo {author} {\bibfnamefont {J.}~\bibnamefont {Lautanen}}, \bibinfo {author} {\bibfnamefont {M.}~\bibnamefont {Honkanen}}, \ and\ \bibinfo {author} {\bibfnamefont {J.}~\bibnamefont {Turunen}},\ }\href@noop {} {\bibfield  {journal} {\bibinfo  {journal} {Optik-International Journal for Light and Electron Optics}\ }\textbf {\bibinfo {volume} {110}},\ \bibinfo {pages} {137} (\bibinfo {year} {1999})}\BibitemShut {NoStop}%
    \bibitem [{\citenamefont {Siviloglou}\ and\ \citenamefont {Christodoulides}(2007)}]{Siviloglou_2007}%
      \BibitemOpen
      \bibfield  {author} {\bibinfo {author} {\bibfnamefont {G.~A.}\ \bibnamefont {Siviloglou}}\ and\ \bibinfo {author} {\bibfnamefont {D.~N.}\ \bibnamefont {Christodoulides}},\ }\href {\doibase 10.1364/OL.32.000979} {\bibfield  {journal} {\bibinfo  {journal} {Opt. Lett.}\ }\textbf {\bibinfo {volume} {32}},\ \bibinfo {pages} {979} (\bibinfo {year} {2007})}\BibitemShut {NoStop}%
    \bibitem [{\citenamefont {Anaya-Contreras}\ \emph {et~al.}(2021)\citenamefont {Anaya-Contreras}, \citenamefont {Zúñiga-Segundo},\ and\ \citenamefont {Moya-Cessa}}]{Arturo}%
      \BibitemOpen
      \bibfield  {author} {\bibinfo {author} {\bibfnamefont {J.~A.}\ \bibnamefont {Anaya-Contreras}}, \bibinfo {author} {\bibfnamefont {A.}~\bibnamefont {Zúñiga-Segundo}}, \ and\ \bibinfo {author} {\bibfnamefont {H.~M.}\ \bibnamefont {Moya-Cessa}},\ }\href {\doibase 10.1364/JOSAA.418533} {\bibfield  {journal} {\bibinfo  {journal} {J. Opt. Soc. Am. A}\ }\textbf {\bibinfo {volume} {38}},\ \bibinfo {pages} {711} (\bibinfo {year} {2021})}\BibitemShut {NoStop}%
    \bibitem [{\citenamefont {Bandres}\ and\ \citenamefont {Guti\'{e}rrez-Vega}(2007)}]{Bandres_2007}%
      \BibitemOpen
      \bibfield  {author} {\bibinfo {author} {\bibfnamefont {M.~A.}\ \bibnamefont {Bandres}}\ and\ \bibinfo {author} {\bibfnamefont {J.~C.}\ \bibnamefont {Guti\'{e}rrez-Vega}},\ }\href {\doibase 10.1364/OE.15.016719} {\bibfield  {journal} {\bibinfo  {journal} {Opt. Express}\ }\textbf {\bibinfo {volume} {15}},\ \bibinfo {pages} {16719} (\bibinfo {year} {2007})}\BibitemShut {NoStop}%
    \bibitem [{\citenamefont {Khonina}(2011)}]{Khonina_2011}%
      \BibitemOpen
      \bibfield  {author} {\bibinfo {author} {\bibfnamefont {S.~N.}\ \bibnamefont {Khonina}},\ }\href {\doibase https://doi.org/10.1016/j.optcom.2011.05.068} {\bibfield  {journal} {\bibinfo  {journal} {Optics Communications}\ }\textbf {\bibinfo {volume} {284}},\ \bibinfo {pages} {4263} (\bibinfo {year} {2011})}\BibitemShut {NoStop}%
    \bibitem [{\citenamefont {Arriz\'{o}n}\ \emph {et~al.}(2018)\citenamefont {Arriz\'{o}n}, \citenamefont {Soto-Eguibar}, \citenamefont {S\'{a}nchez-De-La-Llave},\ and\ \citenamefont {Moya-Cessa}}]{Arrizon_2018}%
      \BibitemOpen
      \bibfield  {author} {\bibinfo {author} {\bibfnamefont {V.}~\bibnamefont {Arriz\'{o}n}}, \bibinfo {author} {\bibfnamefont {F.}~\bibnamefont {Soto-Eguibar}}, \bibinfo {author} {\bibfnamefont {D.}~\bibnamefont {S\'{a}nchez-De-La-Llave}}, \ and\ \bibinfo {author} {\bibfnamefont {H.~M.}\ \bibnamefont {Moya-Cessa}},\ }\href {\doibase 10.1364/OSAC.1.000604} {\bibfield  {journal} {\bibinfo  {journal} {OSA Continuum}\ }\textbf {\bibinfo {volume} {1}},\ \bibinfo {pages} {604} (\bibinfo {year} {2018})}\BibitemShut {NoStop}%
    \bibitem [{\citenamefont {Stoler}(1981)}]{Stoler_1980}%
      \BibitemOpen
      \bibfield  {author} {\bibinfo {author} {\bibfnamefont {D.}~\bibnamefont {Stoler}},\ }\href {\doibase 10.1364/JOSA.71.000334} {\bibfield  {journal} {\bibinfo  {journal} {J. Opt. Soc. Am.}\ }\textbf {\bibinfo {volume} {71}},\ \bibinfo {pages} {334} (\bibinfo {year} {1981})}\BibitemShut {NoStop}%
    \bibitem [{\citenamefont {Zúñiga-Segundo}\ \emph {et~al.}(2014)\citenamefont {Zúñiga-Segundo}, \citenamefont {Rodr\'{i}guez-Lara}, \citenamefont {{Fernández}-{C.}},\ and\ \citenamefont {Moya-Cessa}}]{David}%
      \BibitemOpen
      \bibfield  {author} {\bibinfo {author} {\bibfnamefont {A.}~\bibnamefont {Zúñiga-Segundo}}, \bibinfo {author} {\bibfnamefont {B.~M.}\ \bibnamefont {Rodr\'{i}guez-Lara}}, \bibinfo {author} {\bibfnamefont {D.~J.}\ \bibnamefont {{Fernández}-{C.}}}, \ and\ \bibinfo {author} {\bibfnamefont {H.~M.}\ \bibnamefont {Moya-Cessa}},\ }\href {\doibase 10.1364/OE.22.000987} {\bibfield  {journal} {\bibinfo  {journal} {Opt. Express}\ }\textbf {\bibinfo {volume} {22}},\ \bibinfo {pages} {987} (\bibinfo {year} {2014})}\BibitemShut {NoStop}%
    \bibitem [{\citenamefont {Moya-Cessa}\ and\ \citenamefont {Soto-Eguibar}(2011)}]{Libro}%
      \BibitemOpen
      \bibfield  {author} {\bibinfo {author} {\bibfnamefont {H.~M.}\ \bibnamefont {Moya-Cessa}}\ and\ \bibinfo {author} {\bibfnamefont {F.}~\bibnamefont {Soto-Eguibar}},\ }\href@noop {} {\emph {\bibinfo {title} {Introduction to Quantum Optics}}}\ (\bibinfo {year} {2011})\BibitemShut {NoStop}%
    \bibitem [{\citenamefont {Moya-Cessa}\ \emph {et~al.}(2024)\citenamefont {Moya-Cessa}, \citenamefont {Ramos-Prieto}, \citenamefont {{S\'anchez}-{de-la-Llave}}, \citenamefont {Ru\'{\i}z}, \citenamefont {Arriz\'on},\ and\ \citenamefont {Soto-Eguibar}}]{CRBI}%
      \BibitemOpen
      \bibfield  {author} {\bibinfo {author} {\bibfnamefont {H.~M.}\ \bibnamefont {Moya-Cessa}}, \bibinfo {author} {\bibfnamefont {I.}~\bibnamefont {Ramos-Prieto}}, \bibinfo {author} {\bibfnamefont {D.}~\bibnamefont {{S\'anchez}-{de-la-Llave}}}, \bibinfo {author} {\bibfnamefont {U.}~\bibnamefont {Ru\'{\i}z}}, \bibinfo {author} {\bibfnamefont {V.}~\bibnamefont {Arriz\'on}}, \ and\ \bibinfo {author} {\bibfnamefont {F.}~\bibnamefont {Soto-Eguibar}},\ }\href {\doibase 10.1103/PhysRevA.109.043528} {\bibfield  {journal} {\bibinfo  {journal} {Phys. Rev. A}\ }\textbf {\bibinfo {volume} {109}},\ \bibinfo {pages} {043528} (\bibinfo {year} {2024})}\BibitemShut {NoStop}%
    \bibitem [{\citenamefont {Ramos-Prieto}\ \emph {et~al.}(2024)\citenamefont {Ramos-Prieto}, \citenamefont {{Sánchez}-{de-la-Llave}}, \citenamefont {Ruíz}, \citenamefont {Arrizón}, \citenamefont {Soto-Eguibar},\ and\ \citenamefont {Moya-Cessa}}]{CRBII}%
      \BibitemOpen
      \bibfield  {author} {\bibinfo {author} {\bibfnamefont {I.}~\bibnamefont {Ramos-Prieto}}, \bibinfo {author} {\bibfnamefont {D.}~\bibnamefont {{Sánchez}-{de-la-Llave}}}, \bibinfo {author} {\bibfnamefont {U.}~\bibnamefont {Ruíz}}, \bibinfo {author} {\bibfnamefont {V.}~\bibnamefont {Arrizón}}, \bibinfo {author} {\bibfnamefont {F.}~\bibnamefont {Soto-Eguibar}}, \ and\ \bibinfo {author} {\bibfnamefont {H.}~\bibnamefont {Moya-Cessa}},\ }\href {\doibase https://doi.org/10.1016/j.ijleo.2024.171864} {\bibfield  {journal} {\bibinfo  {journal} {Optik}\ }\textbf {\bibinfo {volume} {309}},\ \bibinfo {pages} {171864} (\bibinfo {year} {2024})}\BibitemShut {NoStop}%
    \bibitem [{\citenamefont {Berry}\ and\ \citenamefont {Balazs}(1979)}]{Berry}%
      \BibitemOpen
      \bibfield  {author} {\bibinfo {author} {\bibfnamefont {M.~V.}\ \bibnamefont {Berry}}\ and\ \bibinfo {author} {\bibfnamefont {N.~L.}\ \bibnamefont {Balazs}},\ }\href {\doibase 10.1119/1.11855} {\bibfield  {journal} {\bibinfo  {journal} {American Journal of Physics}\ }\textbf {\bibinfo {volume} {47}},\ \bibinfo {pages} {264} (\bibinfo {year} {1979})}\BibitemShut {NoStop}%
    \bibitem [{\citenamefont {Eichelkraut}\ \emph {et~al.}(2014)\citenamefont {Eichelkraut}, \citenamefont {Vetter}, \citenamefont {Perez-Leija}, \citenamefont {Moya-Cessa}, \citenamefont {Christodoulides},\ and\ \citenamefont {Szameit}}]{Optica}%
      \BibitemOpen
      \bibfield  {author} {\bibinfo {author} {\bibfnamefont {T.}~\bibnamefont {Eichelkraut}}, \bibinfo {author} {\bibfnamefont {C.}~\bibnamefont {Vetter}}, \bibinfo {author} {\bibfnamefont {A.}~\bibnamefont {Perez-Leija}}, \bibinfo {author} {\bibfnamefont {H.}~\bibnamefont {Moya-Cessa}}, \bibinfo {author} {\bibfnamefont {D.~N.}\ \bibnamefont {Christodoulides}}, \ and\ \bibinfo {author} {\bibfnamefont {A.}~\bibnamefont {Szameit}},\ }\href {\doibase 10.1364/OPTICA.1.000268} {\bibfield  {journal} {\bibinfo  {journal} {Optica}\ }\textbf {\bibinfo {volume} {1}},\ \bibinfo {pages} {268} (\bibinfo {year} {2014})}\BibitemShut {NoStop}%
    \bibitem [{\citenamefont {Dattoli}\ and\ \citenamefont {Torre}(2014)}]{Dattoli}%
      \BibitemOpen
      \bibfield  {author} {\bibinfo {author} {\bibfnamefont {G.}~\bibnamefont {Dattoli}}\ and\ \bibinfo {author} {\bibfnamefont {A.}~\bibnamefont {Torre}},\ }\href {\doibase 10.1364/JOSAB.31.002214} {\bibfield  {journal} {\bibinfo  {journal} {J. Opt. Soc. Am. B}\ }\textbf {\bibinfo {volume} {31}},\ \bibinfo {pages} {2214} (\bibinfo {year} {2014})}\BibitemShut {NoStop}%
    \end{thebibliography}
%
    \end{document}